%% file: main.tex
\documentclass[sigconf]{acmart}

\input{math_commands.tex}

\usepackage[utf8]{inputenc} 
\usepackage[T1]{fontenc}    
\usepackage{url}            
\usepackage{booktabs}       
\usepackage{amsfonts}       
\usepackage{nicefrac}       
\usepackage{microtype}      
\usepackage{xcolor}         
\usepackage{color}
\usepackage{cases}

\usepackage{makecell}
\usepackage{multirow}
\usepackage{booktabs}
\usepackage{enumitem}
\usepackage{listings}
\usepackage{caption}
\usepackage{dsfont}
\usepackage{graphicx}
\usepackage{algorithm}
\usepackage{algorithmic}

\usepackage{subfig}
\usepackage{ulem}
\usepackage{afterpage}
\usepackage{xspace}
\usepackage{adjustbox}
\usepackage{fancybox}

\theoremstyle{plain}

\theoremstyle{definition}

\theoremstyle{remark}

\newcommand{\method}{{HiCom}\xspace}

\AtBeginDocument{%
  \providecommand\BibTeX{{%
    \normalfont B\kern-0.5em{\scshape i\kern-0.25em b}\kern-0.8em\TeX}}}

\setcopyright{acmlicensed}
\copyrightyear{2018}
\acmYear{2018}
\acmDOI{XXXXXXX.XXXXXXX}

\acmConference[Conference acronym 'XX]{Make sure to enter the correct
  conference title from your rights confirmation emai}{June 03--05,
  2018}{Woodstock, NY}
\acmISBN{978-1-4503-XXXX-X/18/06}

\begin{document}

\title{Hierarchical Compression of Text-Rich Graphs via Large Language Models}

\author{Shichang Zhang}
\authornote{Work done while being an intern at Amazon Web Services.}
\affiliation{
  \institution{University of California, Los Angeles}
  \country{}
  \city{}
}
\email{shichang@cs.ucla.edu}

\author{Da Zheng}
\affiliation{
  \institution{Amazon}
  \country{}
  \city{}
}
\email{dzzhen@amazon.com}

\author{Jiani Zhang}
\affiliation{
  \institution{Amazon}
  \country{}
  \city{}
}
\email{zhajiani@amazon.com}

\author{Qi Zhu}
\affiliation{
  \institution{Amazon}
  \country{}
  \city{}
}
\email{qzhuamzn@amazon.com}

\author{Xiang Song}
\affiliation{
  \institution{Amazon}
  \country{}
  \city{}
}
\email{xiangsx@amazon.com}

\author{Soji Adeshina}
\affiliation{
  \institution{Amazon}
  \country{}
  \city{}
}
\email{adesojia@amazon.com}

\author{Christos Faloutsos}
\affiliation{
  \institution{Carnegie Mellon University}
  \institution{Amazon}
  \country{}
  \city{}
}
\email{christos@cs.cmu.edu}

\author{George Karypis}
\affiliation{
  \institution{Amazon}
  \country{}
  \city{}
}
\email{gkarypis@amazon.com}

\author{Yizhou Sun}
\affiliation{
  \institution{University of California, Los Angeles}
  \institution{Amazon}
  \country{}
  \city{}
}
\email{yzsun@cs.ucla.edu}

\renewcommand{\shortauthors}{Zhang, et al.}

\begin{abstract}
\input{sections/abstract}
\end{abstract}



\received{20 February 2007}
\received[revised]{12 March 2009}
\received[accepted]{5 June 2009}

\maketitle

\section{Introduction}\label{sec:introduction}
\input{sections/introduction}

\section{Related Work}\label{sec:related}
\input{sections/related}

\section{Notations and Preliminaries} \label{sec:preliminary}
\input{sections/preliminary}

\section{Method}\label{sec:method}
\input{sections/method}

\section{Experiments} \label{sec:experiment}
\input{sections/experiment}

\section{Conclusion}\label{sec:conclusion}
\input{sections/conclusion}


\bibliographystyle{ACM-Reference-Format}
\bibliography{reference}

\appendix

\section{Appendix}
\input{sections/appendix}

\end{document}

%% file: math_commands.tex
\usepackage{amsmath,amsfonts,bm}

\def\eqref#1{equation~\ref{#1}}

\def\1{\bm{1}}

\def\vs{{\bm{s}}}

\def\vx{{\bm{x}}}

\def\mM{{\bm{M}}}

\def\mS{{\bm{S}}}

\def\mX{{\bm{X}}}

\DeclareMathAlphabet{\mathsfit}{\encodingdefault}{\sfdefault}{m}{sl}
\SetMathAlphabet{\mathsfit}{bold}{\encodingdefault}{\sfdefault}{bx}{n}

\def\gG{{\mathcal{G}}}

\def\V{{\mathcal{V}}}

%% file: sections/abstract.tex
Text-rich graphs, prevalent in data mining contexts like e-commerce and academic graphs, consist of nodes with textual features linked by various relations. Traditional graph machine learning models, such as Graph Neural Networks (GNNs), excel in encoding the graph structural information, but have limited capability in handling rich text on graph nodes. Large Language Models (LLMs), noted for their superior text understanding abilities, offer a solution for processing the text in graphs but face integration challenges due to their limitation for encoding graph structures and their computational complexities when dealing with extensive text in large neighborhoods of interconnected nodes. This paper introduces ``Hierarchical Compression'' (HiCom), a novel method to align the capabilities of LLMs with the structure of text-rich graphs. HiCom processes text in a node's neighborhood in a structured manner by organizing the extensive textual information into a more manageable hierarchy and compressing node text step by step. Therefore, HiCom not only preserves the contextual richness of the text but also addresses the computational challenges of LLMs, which presents an advancement in integrating the text processing power of LLMs with the structural complexities of text-rich graphs. Empirical results show that HiCom can outperform both GNNs and LLM backbones for node classification on e-commerce and citation graphs. HiCom is especially effective for nodes from a dense region in a graph, where it achieves a \textbf{3.48\%} average performance improvement on five datasets while being more efficient than LLM backbones.

%% file: sections/introduction.tex
Text-rich graphs have become prevalent and increasingly important in data mining. These graphs combine textual features with graph structures, offering a unique representation that captures complex entity interactions. For instance, academic graphs represent publications as nodes linked by citations, with paper titles and abstracts providing rich textual node features~\cite{zhang2023effect}. Similarly, e-commerce graphs represent products as nodes connected by customer viewing and purchasing histories, enriched with detailed textual product descriptions~\cite{amazon_data}. These graphs are characterized by both substantial textual information and complex relations. 

\begin{figure}[t]
\begin{center}
\includegraphics[width=\columnwidth]{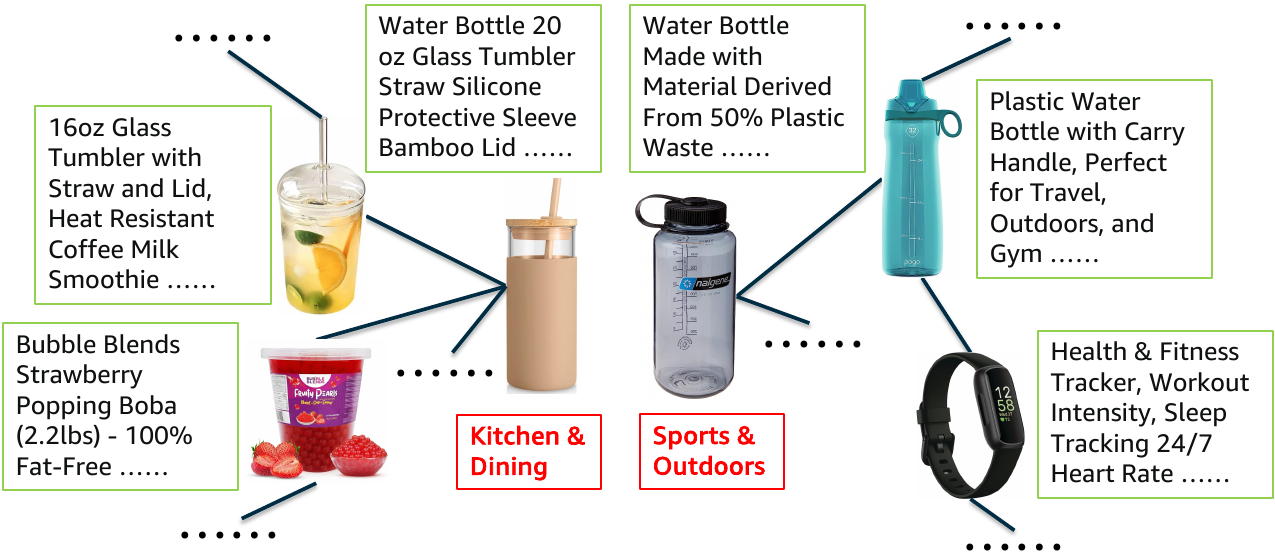}
\end{center}
\caption{Category classification of two water bottles (in the middle) from the Amazon product-co-viewing graph. Their categories (in red) are not clear solely from the product descriptions (in green boxes), but will more likely be correctly classified through the neighborhood context. }
\label{figure:waterbottles}
\Description{Category classification of two water bottles (in the middle) from the Amazon product-co-viewing graph.}
\end{figure}

Studies have pointed out that both textual features and graph structures are crucial for data mining tasks associated with these graphs, e.g., node classification~\cite{ye2021beyond, zhang2021match}. 
Figure~\ref{figure:waterbottles} presents two product-co-viewing graph snippets, demonstrating that product category classifications such as \texttt{Kitchen \& Dining} or \texttt{Sports \& Outdoors} become clearer by considering the surrounding context of neighboring nodes, rather than relying solely on product descriptions.
For instance, the category of a water bottle leans towards \texttt{Kitchen \& Dining} when linked with items such as ``Glass Tumblers'' and ``Strawberry Popping Boba'', suggesting a usage scenario aligned with home and kitchen settings. 
Conversely, a water bottle, with a similar description is more likely to be classified under \texttt{Sports \& Outdoors} if it is often co-viewed with other water bottles ``for travel, outdoors and gym'' as well as ``Health and Fitness Trackers''. Thus, integrating textual descriptions with the relational information from a graph structure captures the comprehensive neighborhood context and enhances predictive accuracy.

To leverage both textual features and graph structures for solving problems on text-rich graphs, Graph Neural Networks (GNNs) have recently gained popularity as they can encode both information via message passing~\cite{gcn, gat, graphsage}. While GNNs are good at capturing structural information and have achieved great results on many benchmarks~\cite{hu2020ogb, morris2020tudataset, sen2008collective}, they are not inherently equipped to handle raw text directly. In the situation of text-rich graphs, the raw text is usually processed by a text encoder, e.g., a language model (LM), to transform it into a feature vector before GNNs can be applied. However, as evidenced by previous studies~\cite{chien2021node, zhu2023touchup}, separately encoding text and performing message-passing is suboptimal in capturing semantic meanings of the neighborhood context and can result in inferior performance. When textual data volume is large and contextually rich, there is a clear demand for strategies that can seamlessly integrate the graph structures with the textual features.

The exceptional ability of large language models (LLMs)~\cite{radford2019language, brown2020language, touvron2023llama, sanh2021multitask, zhang2022opt} in solving language tasks positions them as promising tools for handling text-rich graph problems. Their ability to contextualize text and draw semantic relationships provides potential solutions to address the challenges where GNNs fall short. However, integrating LLMs with text-rich graphs presents significant challenges and remains under-explored. 
One major challenge is that LLMs, originally designed for one-dimensional sequential text, face difficulties when dealing with complex topological structures of graphs. Additionally, LLMs are constrained by a token limit that is inadequate for the dense and abundant text in real-world graphs. While advanced models such as GPT4~\cite{achiam2023gpt} can handle very long inputs, the computational demands are substantial, necessitating extensive hardware resources. Commonly available open-source LLMs face limitations, only able to handle a few thousand tokens~\cite{brown2020language, sanh2021multitask, touvron2023llama, zhang2022opt}, which is insufficient given the extensive text data contained within the neighborhoods of actual graphs.
For example, an e-commerce graph involving sports equipments on Amazon has an average node degree of 8.14 and about 130.93 text tokens per node~\cite{amazon_data}. Processing even a two-hop neighborhood quickly becomes computationally infeasible. 
Therefore, adapting LLMs to handle text-rich graphs necessitates a method that can effectively compress neighborhood text while preserving structural information.

To address the above challenges, we present \textit{Hierarchical Compression} (\method) enabled by LLMs, which can both model graph structures and manage long text. We draw inspiration from the message-passing (MP) algorithm of GNNs~\cite{graphsage}, as well as the transformer compression techniques for handling long documents, especially AutoCompressor~\cite{chevalier2023adapting}.
\method first constructs a hierarchy according to the graph structure for processing a large neighborhood and then compresses the context level by level to avoid input explosion. 
In our experiments, we apply \method to fine-tune an OPT~\cite{zhang2022opt} model to efficiently capture the semantic meaning of rich neighborhoods. The \method-OPT model outperforms both GNNs and the LLM backbone for node classification on five text-rich graphs covering academic and e-commerce applications. We test \method under both normal settings and also a challenging setting focusing on nodes from a dense region in a graph (k-core graph with densely connected nodes). \method achieves an 3.48\% average performance improvement. Meanwhile, \method is more efficient than vanilla LLMs by breaking long inputs and compressing them level-by-level. Our ablation studies show that \method maintains its superior performance in both dense and sparse regions in the graph as well as when the training data is either scarce or sufficient. In summary, the advantages of \method is multifaceted:

\begin{itemize}[leftmargin=12pt]
\item \textbf{Adaptability}: \method crafts LLMs for easy adaptation to graph inputs by transforming the vast context of a node's neighborhood into a more manageable and structured format.
\item \textbf{Effectiveness}: \method outperforms both GNNs and vanilla LLMs for node classification on text-rich graphs.
\item \textbf{Efficiency}: \method introduces memory-efficient implementation techniques that facilitate LLM fine-tuning on graph data and makes \method more scalable than vanilla LLMs.
\end{itemize}

\begin{figure*}[t]
\begin{center}
\includegraphics[width=\textwidth]{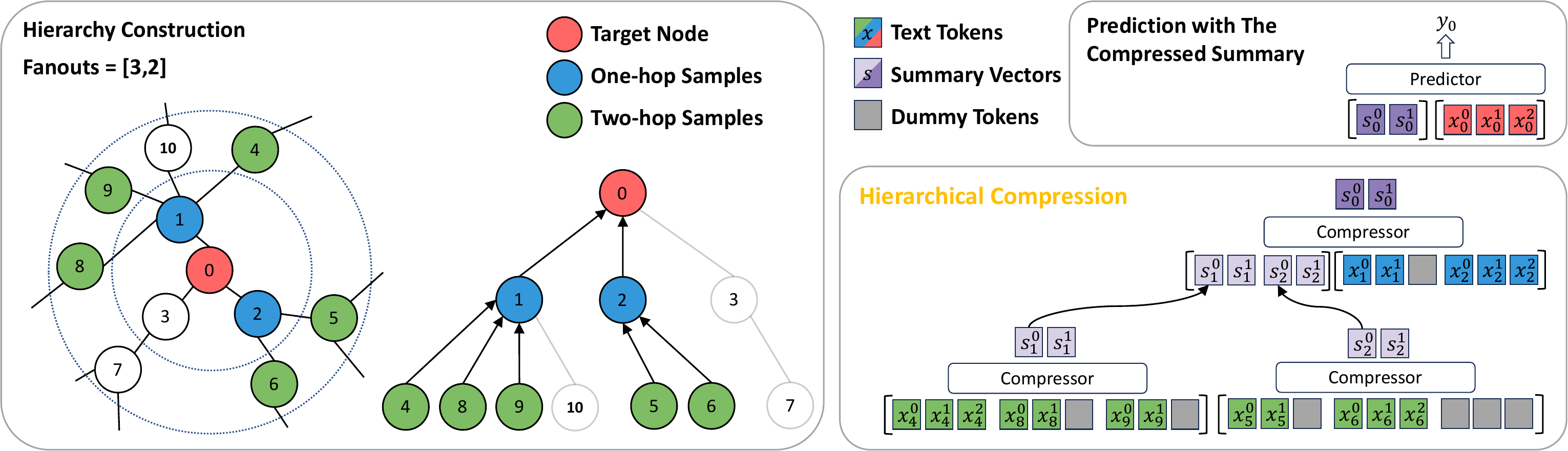}
\end{center}
\caption{An illustration of the hierarchical compression framework with LLMs. The left part corresponds to the hierarchy construction step for a target node $v_0$, with fanouts = [3,2] indicating the budget of nodes to sample in each level. The lower right part shows how the neighborhood context is compressed to a summary vector $s_0$ following the hierarchy. The upper right part shows the final prediction is made with the target node text $x_0$ as input and the summary vector $s_0$.}
\label{figure:framework}
\Description{An illustration of the hierarchical compression framework with LLMs.}
\end{figure*}

%% file: sections/related.tex
\paragraph{Learning on Text-Rich Graphs}
Learning on text-rich graphs requires integration of textual features and graph structures and is gaining increased attention in graph data mining~\cite{zhu2021textgnn, yang2021graphformers, yasunaga2022deep, chien2021node, jin2023patton}. GNNs have shown an exceptional ability to capture structural information and can transform and aggregate node features according to the graph structure via message passing~\cite{gcn, gat, graphsage}. The problem with applying GNNs on text-rich graphs is that they are not inherently equipped to handle raw text. Therefore, GNNs usually need to be combined with a text encoder to transform the text into vector representations, and the encoder may be combined with GNNs through a twin-tower structure~\cite{zhu2021textgnn}, nested layers~\cite{yang2021graphformers}, a cascade structure with joint training~\cite{xie2023graph}, or iterative training via variational inference~\cite{zhao2022learning}. Another approach for learning on text-rich graphs emphasizes pre-training LMs over graph data to preserve relation information and enrich node representations with graph structures~\cite{chien2021node, zhang2020graph, jin2023patton, yasunaga2022deep}. The pre-training objective can be neighborhood prediction~\cite{chien2021node}, feature and structure recovery~\cite{zhang2020graph}, masked modeling at both token-level and document-level~\cite{jin2023patton}, or aligning masked language modeling with knowledge graph reasoning~\cite{yasunaga2022deep}.

\paragraph{Transformer Models for Long Inputs}
Transformers~\cite{vaswani2017attention} have shown their effectiveness in various NLP tasks. Relevant to our research is how transformers handle long inputs~\cite{dai2019transformer, child2019generating, bulatov2022recurrent, choromanski2020rethinking, zheng2022linear, rae2019compressive}. For this line of research, the main challenge is the high complexity and memory demands of attention computation, which limits the input context length of transformers. This issue is particularly pronounced when applying transformers to text-rich graphs, where the context to encode in a neighborhood often far exceeds the model limit. Several methods have been developed for encoding long sequential inputs. Examples include limiting the attention window~\cite{dai2019transformer, child2019generating}, using recurrent memory~\cite{bulatov2022recurrent}, and employing approximate computation of attention~\cite{choromanski2020rethinking, zheng2022linear, rae2019compressive}. The most relevant to this work is AutoCompressor~\cite{chevalier2023adapting}, where a long sequential input is segmented into smaller pieces for step-by-step encoding. We review its technical details in Section~\ref{sec:preliminary}. A detailed survey suggests more methods~\cite{taysurvey}. While these studies are relevant from the model perspective, they do not apply to graph data. Our work bridges this gap, enhancing the ability of transformer-based models to process rich text with graph structures.

\paragraph{Foundation LLMs and Graphs}
Foundation LLMs have shown remarkable versatility and adaptability~\cite{bommasani2021opportunities}. A defining characteristic of these models is their ability to tackle a wide array of tasks through prompting and in-context learning (ICL), without the need for task-specific training. Several works have tried to leverage these foundation LLMs for solving graph tasks through ICL. The ways graph data is handled include using LLMs to summarize node sampled from a neighborhood~\cite{chen2023exploring}, using pure natural language to describe each node’s multi-hop connectivity and meta-features~\cite{fatemi2023talk, ye2023natural, wang2023can}, expanding LLM’s vocabulary by creating a new token for every node~\cite{fatemi2023talk}, and instruction tuning via multiple graph tasks~\cite{tang2023graphgpt}. These explorations are promising but are not directly comparable to this work. 
Our work aims to provide scalable fine-tuned LLMs that can be deployed for various text-rich graphs without concerns of privacy or inference cost. In contrast, many of these works rely on huge black-box LLMs and query those models via API, which causes concerns about scalability, inference cost, and privacy.

%% file: sections/preliminary.tex
This section establishes a common ground in terms of notation and foundational concepts before diving into the core methodologies and results of our study. We summarize the notations that will be used throughout the paper in Table~\ref{tab:notation} and discuss them in the following. In this work, a graph is denoted as $\gG = (\V, \mX)$, where $\V$ represents the set of nodes and $\mX$ represents the text features on nodes. Each node $v_i \in \V$ is associated with a text feature represented in $x_i \in \mX$. We use superscripts to index tokens of text sequence, i.e., $x_i = [x_i^1, ..., x_i^t]$.
For the node classification task, the goal is to predict the label $y_i$ of a node $v_i$ based on its feature and its neighborhood context. As we progress through the paper, further notations will be introduced as required to provide clarity on specific algorithms and methodologies.

\paragraph{Soft Prompts} 
A prompt for LLMs refers to a set of instructions or questions designed to elicit responses or actions from the model, usually in terms of natural text, and is concatenated with the input text data. Soft prompting refers to a technique where instead of providing textual prompts, learnable embeddings are provided as "soft prompts" to guide the model's response~\cite{lester2021power, zhong2021factual, liu2022p}. Similarly, the soft prompts are often concatenated with the tokenized input text data, but they are learnable and will be optimized for the best model response while freezing the LLM parameters. 

\paragraph{AutoCompressor~\cite{chevalier2023adapting}} 
To make LLMs handle long inputs, AutoCompressor applies soft prompts and encodes a long input by breaking it into shorter pieces. AutoCompressor incorporates a series of new learnable tokens $[p^1, \ldots, p^k]$ as soft prompts into LLM's existing vocabulary. These tokens, when appended to the input text, instruct the LLM to output summaries in a vector format. For instance, the input data $[x^{1}, \ldots, x^{t}]$ is appended to include the special tokens, resulting in $[x^{1}, \ldots, x^{t}, p^1, \ldots, p^k]$. This augmented input is processed by the LLM to update the representation of all tokens, including the special tokens. The updated representations of $[p^1, \ldots, p^k]$, denoted as $\vs = [s^{1}, \ldots, s^{k}]$, act as the summary vectors. These summary vectors distill the meaning from the original text input while substantially shortening it to length $k$, where $k << t$. Given such length reduction, a long sequential input can be segmented into smaller pieces and compressed to length $k$ pieces step by step. The learning of these soft prompt tokens slightly alters the standard decoder-based LLMs' next-token prediction objective. A standard objective focuses on maximizing the log probability of the next token conditioned on the previously observed tokens, i.e., $\log p(x^{t+i} | x^{1}, \ldots, x^{t+i-1})$. AutoCompressor soft prompt learning involves substituting context tokens with summary vectors, i.e. changing the objective to $\log p(x^{t+i} | s^{1}, \ldots, s^{k}, x^{t+1}, \ldots, x^{t+i-1})$, where the original context $[x^{1}, \ldots, x^{t}]$ is replaced by $[s^{1}, \ldots, s^{k}]$.

\input{tables/notations}

%% file: tables/notations.tex
\begin{table}[t]
\centering
\caption{Notations}
\resizebox{\columnwidth}{!}{
\begin{tabular}{@{}clll@{}}
\toprule
 Notations & Descriptions \\ \midrule
$\gG = (\V, \mX)$ & A graph with text nodes $\V$ and text features $\mX$. \\
$v_i \in \V$ & Nodes in $\gG$. \\
$x_i \in \mX$ & Raw text on node $v_i$, with $x_i^j$ stands for the $j$-th token. \\
$\vs_i$ & Summary vectors of $v_i$'s neighborhood context. \\
$p$ & The soft prompt, with $p^j$ stands for the $j$-th token. \\
$Comp(\cdot)$ & The LLM-based compressor that maps text (and optional \\ & \quad lower-level summary vectors) to summary vectors. \\
$L$ & Number of hierarchy levels. \\
$[n_1, \cdots, n_L]$ & Budget of nodes to sample for each level (fanouts).  \\
$\mX^C (\tilde \mX^C)$ & Concatenated text ($\tilde \mX^C$ means reshaped). \\
$\mS^C (\tilde \mS^C)$ & Concatenated summary vectors ($(\tilde \mS^C)$ means reshaped). \\
\bottomrule
\end{tabular}
}
\label{tab:notation}
\end{table}

%% file: sections/method.tex
Hierarchical compression (\method) is a framework designed to integrate LLMs with text-rich graphs. Its motivation lies in leveraging the strong text-understanding ability of LLMs to extract contextual information from an extensive neighborhood. The primary challenge is to process the complex graph structure and the abundant text in that neighborhood, 
since LLMs are originally designed for one-dimensional sequential data only and have a limited input length that is often much shorter than the total amount of neighborhood context. Therefore, integrating LLMs for text-rich graphs necessitates a method that can effectively compress neighborhood text while still preserving structural information. \method achieves the goal by compressing the neighborhood in a structured manner. 
We construct a hierarchy of feature aggregation that corresponds to different hops from the center node, especially building the iterative process of aggregating textual features from a node's immediate neighbors to capture local graph structures, and then progressively expanding this aggregation to include distant neighbors. 
Such design enables efficient usage of the graph structure and avoids input explosion that would occur if one were to simply concatenating all node attributes in sequence for encoding a large neighborhood. \method adapts a pre-trained LLM as the backbone and adds additional learnable vocabularies and prediction layers to make the model work as a graph compressor and a predictor. All modules are jointly fine-tuned with parameter-efficient fine-tuning (PEFT) to solve graph tasks, e.g., node classification. 

In the following paragraphs, we begin by demonstrating the workflow of the \method framework with an illustrative example, and then provide details about the compressor, the predictor, and the computational complexity associated with \method. Furthermore, we discuss techniques to improve the efficiency and effectiveness of \method and examine its relationship with GNN message passing. 

\subsection{The HiCom Framework}
\paragraph{The Workflow}
We illustrate how \method works through a node classification example shown in Figure \ref{figure:framework}. This illustration is for classifying the target node $v_0$. Initially, \method constructs a hierarchy with $L$ levels for gathering context information in the neighborhood of $v_0$, with $v_0$ in the last ($L$-th) level (left figure). The hierarchy is formed by sampling multi-hop neighbors of $v_0$, and the size of each level is specified by the ``fanouts'' parameters, with the $l$-th entry for level $l$. For instance, this example illustrates the construction of a two-level hierarchy with fanouts set to [3,2], which means two one-hop neighbors (e.g., $v_1$ and $v_2$) of $v_0$ and three two-hop neighbors for each one-hop neighbor (e.g., $v_4$, $v_8$, and $v_9$ for $v_1$; $v_5$ and $v_6$ for $v_2$) are sampled. (Note that while three neighbors were intended for $v_2$ following the fanouts, only two are included, as those are all $v_2$ has. This aspect introduces an implementation challenge that will be elaborated in Section~\ref{subsec:implementation}).
The constructed hierarchy is represented as a tree structure. Considering the potentially extensive context within this hierarchy, the compression phase is the next step, as shown in the bottom right figure. The compressor concatenates the text tokens in each local neighborhood and compresses them into summary vectors. The summary vectors capture essential contextual information while significantly reducing the input length. In this example, an input in the lower hierarchy (green tokens) is compressed from length nine into length two. (Note that dummy tokens might be needed as placeholders for parallelization, a topic will be elaborated in Section~\ref{subsec:implementation}). 
After two sets of summary vectors, $\vs_1$ and $\vs_2$, have been produced, they are further compressed along with the text on $v_1$ and $v_2$ to produce a comprehensive summary of $v_0$'s neighborhood. The final summary vectors for $v_0$ become $s_0$. Lastly, the predictor combines $s_0$ with the raw text of the target node, denoted as $x_0$, to predict the label $y_0$, as illustrated in the upper right figure.

\paragraph{The Compressor} The compressor processes the text from neighboring nodes in a level-by-level manner. For the text data in the $\ell$-th level, the compressor employs soft prompting as in AutoCompressor (reviewed in Section~\ref{sec:preliminary}) to instruct LLMs to compress the input text into summary vectors. $k$ soft prompt tokens are added to LLM's existing vocabulary to compress inputs with length $t$ to summary vectors with length $k$. We denote this compression operation as $\vs = Comp(x)$, omitting the token-level detail. An example is depicted in Figure~\ref{figure:framework} with $k$ equals 2 for visualization purposes, and we set $k$ to be 50 in our experiments. 
To compress input from different levels hierarchically, the summary vectors from $\ell$-th level nodes in the computation graph are consumed by the $(\ell+1)$-th level, being prepended to their embedding inputs. Thus, the summary from level $\ell$ can be carried to level $(\ell+1)$ for further compression. For example, the compression of $v_0$'s neighborhood in Figure~\ref{figure:framework} is $\vs_0 = Comp([\vs_{1}, \vs_{2}, x_{1}, x_{2}])$.  $\vs_0$ will thus contain not only information from $x_{1}, x_{2}$, but also information from $x_{4}, x_{8}, x_{9}, x_{5}$, and $x_{6}$ through hierarchical compression.

\paragraph{The Predictor}
When using the LLM as a predictor, we append and tune an extra prediction layer at the end of the model. The parameters in this layer are randomly initialized and tuned together with the LLM backbone. Similarly to the compressor operation, the predictor can also take in the combined summary vectors and text tokens, and therefore make predictions with both the node text and the contextual information. For example, to make a prediction of $v_0$ in the illustration example, $[\vs_{0}, x_{0}]$ is fed to the predictor as input, which concatenates the neighborhood summary vectors and the feature information of node $v_0$.

\paragraph{Computational Complexity}
One of the advantages of \method is its efficiency for processing rich neighborhoods compared to the LLM backbone. This efficiency stems from the fact that the computational complexity of LLM forward passes grows quadratically with the input length. Specifically, if we were to concatenate sampled neighbors directly to the target to form an input of length $n$, the resulting time complexity would be $O(n^2)$. However, by dividing the input into smaller segments of lengths $n_1, n_2, \ldots, n_t$ (s.t. $n = \sum_{i=1}^{t} n_i$) and compressing them hierarchically, the overall time complexity is reduced to $O(\sum_{i=1}^{t} n_i^2)$. \method results in a lower computational cost for processing long inputs, and its advantage is more prominent for longer inputs.

\subsection{Techniques for Efficiency and Effectiveness}\label{subsec:implementation}

In this section, we discuss some techniques to improve the efficiency and effectiveness of \method. The efficiency perspective is important as LLMs can have multi-billion parameters with considerable GPU memory requirements. We thus introduce implementation techniques for batch processing of neighborhoods with different sizes 
and reducing the memory consumption of unnecessary padding. These techniques are not restricted to the \method algorithm and can be used for general LLMs to optimize GPU utilization on graph data.
From the effectiveness perspective, we discuss the idea of summary accumulation which shares a similar idea to skip connection.

\subsubsection{Batch Processing of Neighborhoods with Different Sizes}
Algorithm~\ref{alg:hc_single} presents a straightforward version of \method for individual nodes, offering a fundamental understanding of the algorithm. In practice, batch processing is essential for maximizing the efficiency of GPU utilization, but it also introduces specific challenges.
Since LLMs are innately designed for sequential data, the challenge in adapting \method with LLMs for batch processing primarily arises from the variable-length inputs produced by graph data, which complicates the batching of data instances. To address these challenges, a more sophisticated padding strategy is needed to uniform both the length of the text sequence on each node and the number of neighbors of each node for all nodes in the same level of the hierarchy. Algorithm~\ref{alg:hc_batch} outlines \method for processing nodes in batches. We discuss the algorithm in detail in Appendix~\ref{app:batch_process}.

\begin{algorithm}[t]
  \caption{\method for processing individual nodes}
  \label{alg:hc_single}
\begin{algorithmic}[1]
  \STATE {\bfseries Input:} $\gG = (\V, \mX)$, fanouts=[$n_1, \cdots, n_L $], $Comp(\cdot)$, a target node $v \in \V$ 
  \STATE {\bfseries Output:} Summary vectors $s_v$ of node $v$ 
  \STATE $\V_{B_L} = \{v\}$ // a batch with a single node, i.e., index $B_L = v$
  \FOR{$l=L$ {\bfseries to} $1$}
    \STATE $\V_{B_{l-1}} = $ Sample up to $n_l$ neighbors of $v$, $\forall v \in V_{B_{l}}$ 
    \STATE $\mX_{B_{l-1}} = $ Collect $x_{v}$, $\forall v \in V_{B_{l-1}}$
  \ENDFOR
  \STATE $\mS_{B_{0}}$ = []
  \FOR{$l=1$ {\bfseries to} $L$}
    \STATE $\mX_{B_{l}}^C$ = Concat($\mX_{B_{l-1}}$)
    \STATE $\mS_{B_{l}}^C$ = Concat($\mS_{B_{l-1}}$)
    \STATE $\mS_{B_{l}}$ = $Comp([\mS_{B_{l}}^C, \mX_{B_{l}}^C])$
  \ENDFOR
  \RETURN $\mS_{B_{L}}$ // which is $s_v$
\end{algorithmic}
\end{algorithm}

\begin{algorithm}[t]
  \caption{\method for processing nodes in batches.}
  \label{alg:hc_batch}
\begin{algorithmic}[1]
  \STATE {\bfseries Input:} $\gG = (\V, \mX)$, fanouts=[$n_1, \cdots, n_L $], $Comp(\cdot)$, a batch of target nodes $\V_{B} \in \V$ indexed by $B$. The sequence length is $t$, and the number of summary tokens is $k$.
  \STATE {\bfseries Output:} Summary vectors $\mS_{B}$ of nodes $\V_{B}$ 
  \STATE $\V_{B_L} = \V_{B}$
  \FOR{$l=L$ {\bfseries to} $1$}
    \STATE $\V_{B_{l-1}} = $ Sample up to $n_l$ neighbors of $v$, $\forall v \in V_{B_{l}}$ 
    \STATE $\mX_{B_{l-1}} = $ Collect $x_{v}$ and pad to length $t$, $\forall v \in V_{B_{l-1}}$
  \ENDFOR
  \STATE $\mS_{B_{0}}$ = A placeholder of all zeros with shape $[|V_{B_{0}}|, k]$
  \FOR{$l=1$ {\bfseries to} $L$}
    \STATE $\mX_{B_{l}}^C$ = A placeholder of all zeros with shape $[|V_{B_{l}}|, n_l, t]$
    \STATE $\mS_{B_{l}}^C$ = A placeholder of all zeros with shape $[|V_{B_{l}}|, n_l, k]$
    \STATE $\mX_{B_{l}}^C[{B_{l-1}}]$ = $\mX_{B_{l-1}}$
    \STATE $\mS_{B_{l}}^C[{B_{l-1}}]$ = $\mS_{B_{l-1}}$
    \STATE $\tilde\mX_{B_{l}}^C$ = Reshape $\mX_{B_{l}}^C$ to $[|V_{B_{l}}| \times n_l, t]$
    \STATE $\tilde\mS_{B_{l}}^C$ = Reshape $\mS_{B_{l}}^C$ to $[|V_{B_{l}}| \times n_l, k]$
    \STATE $\mS_{B_{l}}$ = $Comp([\tilde\mS_{B_{l}}^C, \tilde\mX_{B_{l}}^C])$
  \ENDFOR
  \RETURN $\mS_{B_{L}}$ // which is $\mS_{B}$
\end{algorithmic}
\end{algorithm}

\subsubsection{Pre-tokenization and Rearrange and Trim}
Another effective implementation technique employed in the \method framework to reduce the computational overhead associated with LLMs is pre-tokenization, which involves pre-processing the text data into tokens and padding them with dummy tokens to ensure uniformity in sequence lengths before the training starts. Pre-tokenization saves the redundant time cost of raw text encoding when a node belongs to multiple neighborhoods and is thus involved in multiple computations. However, one inefficiency that arises from pre-tokenization is the presence of dummy tokens within the concatenated sequences for compression, and batching further complicates this problem. For instance, in the example in Figure~\ref{figure:framework}, when sequences [$x_4$, $x_8$, $x_9$] and [$x_5$, $x_6$] are batched into a $2 \times 9$ tensor, only 7 tokens in each sequence are meaningful, with the remainder being dummy tokens but appear in different places of the tensor.

To optimize memory usage and computation time, especially when processing batched sequences, a technique to \textit{rearrange and trim} the sequences is employed. This technique pushes all dummy tokens to one end of the sequence, allowing them to be trimmed off, thereby conserving memory and speeding up computation. In the same example, the $2 \times 9$ tensor can be rearranged and trimmed to $2 \times 7$ to save computation. This rearrange-and-trim method optimizes the efficiency of the \method framework. It is particularly effective when node text has diverse lengths or when the fanout parameter is large. We show a pseudocode of this function in Appendix~\ref{app:rearrange_trim}.

\subsubsection{Summary Accumulation}
To include more explicit context information during both compression and prediction, summary vectors from the previous levels in the hierarchy can be accumulated to enrich the input. For the same example of predicting $v_0$, the input can become $[\vs_{1}, \vs_{2}, \vs_{0}, x_{0}]$ by accumulating $\vs_{1}$, $\vs_{2}$, and $\vs_{0}$ together as an enriched summary. Similarly for compression, if the hierarchy has three levels, summaries from level one can be accumulated to enrich compression in level three. Such operation is referred to as \textit{summary accumulation} in the AutoCompressor model. We discuss its relationship to skip connection and multi-hop filter matrix in GNN message passing (MP) in Section~\ref{subsec:connection} and demonstrate its usefulness in enhancing the model performance in Section~\ref{subsec:abalation}.

\subsection{Connection to Message Passing}~\label{subsec:connection}

This section compares \method with LLMs to GNN MP, highlighting their similarities and distinctions. For similarities, both \method and MP focus on aggregating contextual information from a neighborhood, and they both operate across multi-hop neighbors to gather information level by level. However, there are three critical differences between them. 

Firstly, the situations where these two methods can apply are different. GNN MP is not designed to handle raw text, and the messages passed along have to be feature vectors. In text-rich graphs, this necessitates an additional step of feature extraction from raw text, which can potentially be a bottleneck if the semantic meanings cannot be well preserved by the encoder. Using LLM as encoders or even co-training the encoder with GNNs was shown to be suboptimal~\cite{xie2023graph}, and we will also verify this in our experiments. \method, on the other hand, can be trained end-to-end with raw text and graph as inputs to the final labels as outputs, which is more aligned with the common practice of deep learning models~\cite{lecun2015deep}. 

Secondly, the ways the information is aggregated are different. MP involves an aggregation function to combine all messages passed to a node, usually mean or sum~\cite{gcn, graphsage}. These aggregation functions treat all messages uniformly, without considering their potential influences on each other. Although some GNNs, like GAT~\cite{gat} and GaAN~\cite{zhang2018gaan}, employ the attention mechanism to assign varying weights during message aggregation, the weight is assigned at the message level rather than the token level, making the aggregation still limited in capturing semantic meanings.  In contrast, \method concatenates text sequences on several nodes from a local neighborhood and employs an LLM-based compressor to summarize them, which allows all the words that appear in the neighborhood to interact with each other through the model forward pass and potentially preserves more semantic information. 

Thirdly, the third distinction lies in the utilization of the model-preserved knowledge. GNNs are often trained from scratch and thus have no access to prior knowledge beyond the downstream dataset. Even though there are some GNN pre-training methods, the pre-trained GNN focuses on learning graph structural knowledge~\cite{qiu2020gcc} or domain-specific knowledge~\cite{hu2020gpt, hu2019strategies}. So far, we have not seen GNN pre-training methods that can be as effective as LLM pre-training for memorizing general knowledge extracted from a large corpus. The general knowledge encapsulated in LLMs can be leveraged in \method for generating summary vectors and making predictions, thereby enriching the model's ability to process text data.

Furthermore, the summary accumulation operation used in \method shares the same idea of skip connection~\cite{xu2018representation} and the combination of filter matrices~\cite{frasca2020sign} used in MP. The summary accumulation brings summary from a lower level in the hierarchy to a higher level to make better predictions, which creates a shortcut for information flow across levels, like the skip connection. Similarly, studies of GNN architectures found that combining different filter matrices to pass messages can improve GNN performance, where each filter is responsible for passing messages from a different hop. From this perspective, summary accumulation achieves the same result.

%% file: sections/experiment.tex
\subsection{Datasets}
\input{tables/dataset}

We evaluate our method on two types of text-rich graphs from different domains: the MAPLE academic citation graphs and the graph of Amazon products. Dataset statistics are presented in Table~\ref{tab:dataset}.

\noindent \textbf{MAPLE}~\cite{zhang2023effect}: Metadata-Aware Paper colLEction (MAPLE) is a benchmarking dataset for scientific literature tagging constructed from the Microsoft Academic Graph~\cite{sinha2015overview}. MAPLE contains graphs where papers serve as nodes and citations as edges. Each graph is specific to a specific scientific field, such as mathematics, with nodes representing papers and their corresponding labels indicating subfields, like algebra. This structure sets up a multi-label node classification task. In our experiments, we focus on three specific subfield graphs, namely Mathematics, Economics, and Geology.

\noindent \textbf{Amazon Products}~\cite{amazon_data}: The Amazon products dataset comprises items available on Amazon with detailed descriptions. These items are structured into a graph where each item represents a node. Edges are established between nodes (items) that are co-viewed by users. Furthermore, items are labeled with various product categories, setting the definition for a multi-label node classification task. Given the extensive size of the full graph, which contains over nine million items, our analysis focuses on product graphs from three subdomains, i.e., Sports and Cloth. 

\input{tables/nc}

\subsection{Baselines}
We consider three types of methods. First, we evaluate GNN models that incorporate LM-encoded features. Specifically, we encode the node text using different LMs. Then we feed these LM-encoded features into a GraphSAGE model~\cite{graphsage} for node classification. The LMs we consider include non-fine-tuned BERT~\cite{devlin2018bert}, fine-tuned BERT on the same node classification task~\cite{ioannidis2022efficient}, non-fine-tuned OPT~\cite{zhang2022opt}, GIANT~\cite{chien2021node}, (an XR-Transformer~\cite{zhang2021fast} after self-supervised fine-tuning with graph information), and Patton~\cite{jin2023patton}. For Patton, we use its available fine-tuned model checkpoints on the MAPLE and Amazon Products. These models are denoted as \emph{BERT-GNN}, \emph{BERT-FT-GNN}, \emph{OPT-GNN}, \emph{GIANT-GNN}, and \emph{Patton-GNN}. Second, we consider the \emph{GLEM}~\cite{zhao2022learning} method that iteratively trains a GNN and an LM. Third, we directly fine-tune LLMs on the node text for classification. We also leverage graph context by dynamically sampling and concatenating neighbor text to the node text, which we refer to as LLM neighbor concatenation (NConcat). The LLM we use for this type of approach is the pre-trained OPT model, which is the same LLM used in our proposed \method framework, and the baselines are denoted as \emph{OPT} and \emph{OPT-NConcat}.

\subsection{Experiment Settings}
\paragraph{Data processing.} 
Our main goal is to evaluate the challenging cases where nodes have rich neighborhood information, analogous to long documents in NLP research.
Therefore, we consider two different experiment settings of where the nodes can be sampled from, \textit{dense regions} and \textit{all regions}. For the dense regions, we mimic the challenging long document setting to focus on nodes from the dense region of a graph, where nodes possess rich neighborhood information. In particular, we consider nodes belonging to the k-core graph. 
In the literature, the Amazon product graphs are usually evaluated by taking their 5-core. To make the case more challenging, we consider 8-core for the dense region setting. For all regions, nodes are sampled uniformly randomly, which is more inclusive but not the best setting to test models' ability to handle rich inputs. 

\paragraph{Data splits.} We randomly pick 20 nodes per class as the training set for each graph and sample 1,000 nodes from the rest as the validation set and sample up to 10,000 nodes from the rest as the test set. We evaluate experiments with the F1 score considering the imbalance. We set the number of summary vectors, $k$, to be 50. Given the difference in graph characteristics, we experiment with a few different hyperparameters, e.g. fanouts, to make sure the model produces the best result. We do such hyperparameter tuning for our method as well as the baselines. In Table~\ref{tab:nc_kcore}, we report the test results with the best hyperparameter selected on the validation set. The full results of all methods with different hyperparameters can be found in Table~\ref{tab:nc_detailed} in Appendix~\ref{app:result_detail}. Details of hyperparameters and implementation are shown in Appendix~\ref{app:setting}.

\subsection{Main Results}

The node classification results for the dense regions are shown in Table~\ref{tab:nc_kcore}. We see that our \method-OPT outperforms all baselines on all datasets. We also make some observations and provide related discussions below. 

\paragraph{GNN as Backbone vs. LM as Backbone}
Our first observation is based on the comparison of models from two different categories, e.g., GNN as the backbone (the first multi-row, LM embeddings + GNN) and LM as the backbone (the last two multi-rows, LM fine-tuning and ours). We exclude GLEM from this discussion as it involves in fine-tuning both models for predictions. We observe that all three LM-based methods can provide reasonable performance on these graph datasets, outperforming most GNNs on most datasets, especially for the MAPLE graphs. This observation connects to the reasons we discussed in Section~\ref{subsec:connection}. First, LMs can directly take raw text as input and model token-level interactions through attention, whereas GNNs can only work with encoded embeddings and only model node-level interactions. Secondly, The LM-preserved knowledge from pre-training can be utilized for context understanding and prediction, whereas GNNs do not have preserved knowledge. The second point is more significant on the MAPLE graphs, as these graphs contain academic papers as nodes, which are likely included in the pre-train data of the LM. In contrast, the Amazon data, which consists of diverse and less standardized item descriptions, is less likely to be included in the training data (might be partially included via the Common Crawl data). As a result, fine-tuning only on the text of the target nodes (OPT) has good performance that is close to using neighbor text (OPT-NConcat). 

\paragraph{LM Fine-tuning vs. \method}
Our second observation is the monotone performance improvement from fine-tuning pure LM (OPT) to LM + Neighbors (OPT-NConcat) to \method (\method-OPT). This trend is observed on all datasets for both the dense regions (Table~\ref{tab:nc_kcore}) and all regions (Table~\ref{tab:nc_all}). The improvement from LM to LM + Neighbors is intuitive because more information are added for the LM to process to help with the prediction. However, doing neighbor concatenation also has two bottlenecks. One is that the number of neighbors that can be included is low (< 4 for a 2048 window size of OPT with 512 tokens per node). The other is that the hierarchy in the graph structure is not modeled due to simple concatenation. These are exactly the two bottlenecks that \method overcomes. With many more multi-hop incorporated following a hierarchy, the \method performance further improves over fine-tuning LM + Neighbors.

\begin{figure}[t]
\begin{center}
\includegraphics[width=\columnwidth]{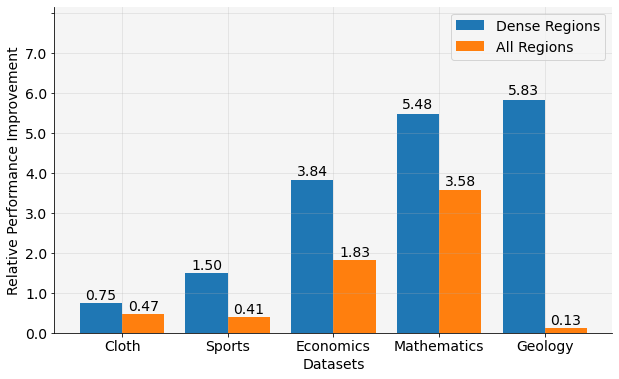}
\end{center}
\caption{\emph{\method always gains}. Relative performance improvement for \method-OPT over the second-best method on nodes in dense regions vs. all regions.}
\label{figure:dense_vs_all}
\Description{Relative performance improvement for \method-OPT over the second-best method on nodes in dense regions vs. all regions.}
\end{figure}

\paragraph{Dense Regions vs. All Regions}
Our results in Table~\ref{tab:nc_kcore} demonstrate that \method-OPT can outperform baselines for nodes in dense regions, especially on graphs with high average degrees. We also evaluate the performance of \method-OPT vs. baselines on randomly selected nodes from all regions. To compare these two cases, we consider the relative performance improvement for \method-OPT over the second-best method. Results for nodes in the dense region vs. nodes in all regions are shown in Figure~\ref{figure:dense_vs_all}. We observe that the performance enhancement of \method-OPT is less significant when applied to nodes in all regions. This observation is expected, as when selecting nodes from all regions, many of them can have only a few neighbors, and hierarchical compression becomes unnecessary. In reality, we observe that a large portion of nodes belong to the dense region, and our method will perform better when the data gets richer with more connections. The detailed experiment setting and performance results on nodes in all regions can be found in Appendix~\ref{app:result_all}.

\paragraph{\method Performance vs. Node Degrees}
We found the \method performance improvements are quite consistent with the average node degrees of the graphs. When we consider the performance improvement of \method-OPT over the second-best model on each dataset in terms of percentage, we get [0.75\%, 1.49\%, 3.83\%, 5.48\%, 5.83\%] improvement in the order of columns presented in the table. This gives a 3.48\% average performance improvement. Also, these numbers are positively correlated with the average node degree reported in Table~\ref{tab:dataset}, with a Pearson correlation coefficient equal to 0.667. This is expected as we discussed above.

\subsection{Ablation Studies} \label{subsec:abalation}
We conduct a series of ablation studies to understand the effectiveness of each module of the proposed \method method and quantify the advantage of \method under different situations.

\input{tables/acc_ablation}

\paragraph{Hierarchy and Summary Accumulation}
We study the significance of two key designs of \method, the hierarchy constructed following the graph structure and the summary accumulation. We saw in the main results that \method-OPT can outperform OPT-NConcat by significant margins. The main reason is that \method-OPT can compress much more neighbor information than a vanilla OPT with limited input length. Here we further show that the hierarchy, i.e., how the graph structure information is used to determine compression order, is also important for \method performance enhancement. We demonstrate this point by considering a case of compressing nodes sampled from the neighborhood in a random order as if these nodes form a set without any graph structure. This ablation can also be seen as applying the OPT-based AutoCompressor directly to a long sequence generated by neighbor concatenation. The difference between this setting and OPT-NConcat is that the number of neighbors in the sequence can be much larger. The result of this ablation study is shown in Table~\ref{tab:acc_ablation}. We see that the performance of \method-OPT without a hierarchy drops significantly compared to the complete version demonstrating the significance of the compression hierarchy. Nonetheless, the ablation performance is still better than OPT-NConcat, which shows the power of compression. Another key design of \method is summary accumulation, which, as we discussed above, acts similarly as skip connections or multi-hop filter matrices to enhance model performance. We also perform an ablation study on it and show results in Table~\ref{tab:acc_ablation}. We observe that removing summary accumulation causes a slight performance drop smaller than removing hierarchy. \method remains the best method on four out of six datasets even with these two ablations.

\begin{figure}[t]
\begin{center}
\includegraphics[width=\columnwidth]{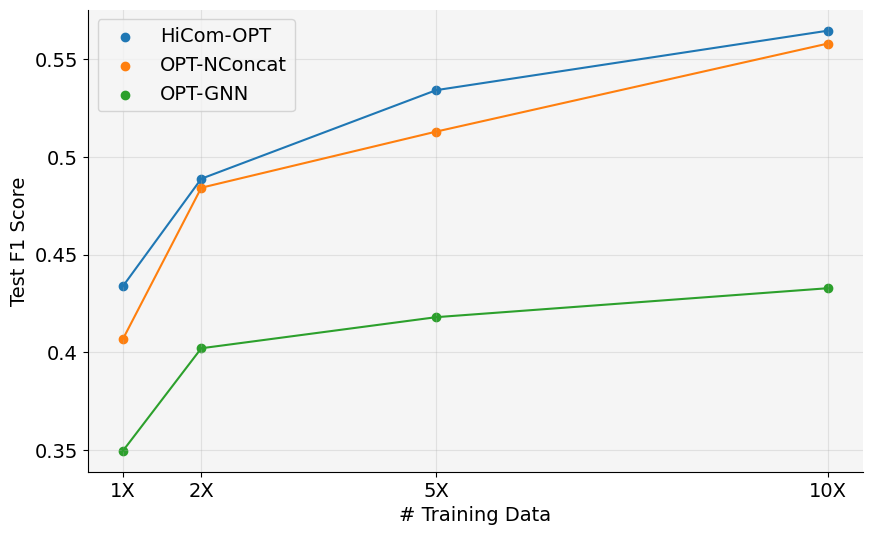}
\end{center}
\caption{\emph{\method wins}. Method performance on the Geology dataset with the training set in different sizes.}
\label{figure:increase_train_data_geology}
\Description{Method performance on the Geology dataset with the training set in different sizes.}
\end{figure}

\paragraph{Increasing Training Data}
In our main experiment, we used a training set comprising 20 samples per class, adhering to the principle that foundational models like LLMs should demonstrate effectiveness with minimal fine-tuning. To further explore the model's robustness and performance, we undertake an ablation study to gradually increase the training set size. Specifically, we expand the training set to be 2, 5, and 10 times larger than the original training set of 20 samples per class. Considering the time-intensive nature of tuning the LLM with a dataset that is 10 times larger than before, we select the Geology graph for this study. The results of these expanded training sets are illustrated in Figure~\ref{figure:increase_train_data_geology}. Our observations indicate that even with the enlarged training set, \method-OPT maintains superior performance compared to the two strong baselines, OPT-NConcat and OPT-GNN. This outcome underscores the efficacy of \method on larger training datasets while still achieving notable performance improvements. The detailed experiment setting and a similar figure for the Sports graph can be found in Appendix~\ref{app:increase_training}.

\subsection{Run Time Comparison}
We discussed that one advantage of \method is its efficiency. To illustrate this, we compare its run time to the backbone OPT with neighbor concatenation (OPT-NConcat). We show their training time for one epoch in Table~\ref{tab:time}. Notably, for a comparable number of neighbors, \method-OPT with fanouts [2,2] only takes $31.24 / 55.46 \approx 56\%$ 
of the time required by OPT-NConcat with 4 neighbors, and the F1-score of \method-OPT is clearly better. Increasing the number of neighbors for compression in \method, e.g. with fanouts equal [2, 8], results in longer training times but also enhanced performance. For OPT-NConcat, however, increasing the number of neighbors is not meaningful as the sequence will be longer than the input limit and get truncated. Additionally, we also verify the effectiveness of the ``rearrange-and-trim'' operation introduced in Section~\ref{subsec:implementation}. Omitting this operation increases the runtime of \method-OPT with fanouts [2,2] from 31.24 to 35.18 minutes, underscoring its significance in boosting efficiency.

\input{tables/time}

%% file: tables/dataset.tex
\begin{table}[t]
    \centering
    \caption{Dataset Statistics. ``Avg. D'' means the average node degree.  ``Avg. T'' means the average number of tokens on each node, estimated by 1 token $= 3/4$ words.}\label{tab:dataset}
    \resizebox{\columnwidth}{!}{
    \begin{tabular}{c|ccccccc}
    \hline
         Dataset & \#Nodes  & \#Edges & \#Classes & Avg. D & Avg. T \\
    \hline
    Cloth      & 469,274 & 2,578,746 & 18 & 5.50  & 173.75 \\
    Sports     & 293,712 & 2,390,076 & 23 & 8.14  & 130.93 \\
    Economics   & 161,727 & 1,530,898 & 40 & 5.45  & 168.48 \\ 
    Mathematics & 437,685 & 2,385,322 & 20 & 9.47  & 198.58 \\ 
    Geology     & 376,503 & 7,622,754 & 18 & 20.25 & 343.89 \\ 
    \bottomrule
    \end{tabular}
    }
\end{table}

%% file: tables/nc.tex
\begin{table*}[thb]
\centering
\caption{\underline{\method wins}: Node classification F1 score on Amazon product graphs and MAPLE citation graphs. For each dataset, the best result is in \textbf{bold} and the second best is \emph{underlined}.}
\begin{tabular}{llccccc}
\toprule
Type & Method & \textbf{\quad Cloth\quad} & \textbf{\quad Sports\quad} & \textbf{Economics} & \textbf{Mathematics} & \textbf{Geology} \\ \midrule
\multirow{5}{*}{LM Embedings + GNN} & BERT-GNN    & 0.6362 & 0.6727 & 0.1822 & 0.2693 & 0.3667 \\
 & BERT-FT-GNN & 0.6457 & 0.6804 & 0.1826 & 0.2861 & 0.3633 \\
 & GIANT-GNN   & 0.6471 & 0.7201 & 0.1651 & 0.2421 & 0.3120 \\
 & Patton-GNN  & \emph{0.6831} & 0.7015 & 0.1664 & 0.2599 & 0.3375 \\
 & OPT-GNN & 0.6484 & 0.7255 & 0.1747 & 0.2713 & 0.3660 \\ \midrule
LM \& GNN Iter Training & GLEM & 0.6743 & \emph{0.7413} & 0.2286 & 0.2834 & 0.4100 \\ \midrule
\multirow{2}{*}{LM Fine-tuning} & OPT & 0.6541 & 0.6005 & 0.2324 & 0.2889 & 0.4206 \\
 & OPT-NConcat & 0.6798 & 0.6728 & \emph{0.2372} & \emph{0.2903} & \emph{0.4218} \\ \midrule
 Ours & \method-OPT & \textbf{0.6882} & \textbf{0.7524} & \textbf{0.2463} & \textbf{0.3062} & \textbf{0.4464} \\
\bottomrule
\end{tabular}
\label{tab:nc_kcore}
\end{table*}

%% file: tables/acc_ablation.tex
\begin{table*}[thb]
\centering
\caption{Node classification F1 score. Ablation on hierarchy (graph structure) and summary accumulation. For each ablation, results with $^*$ mean that \method-OPT with ablation still performs the best on that dataset.}
\begin{tabular}{lccccc}
\toprule
Method & \textbf{\quad Cloth\quad} & \textbf{\quad Sports\quad} & \textbf{Economics} & \textbf{Mathematics} & \textbf{Geology} \\ \midrule
OPT-NConcat    & 0.6798 & 0.6728 & 0.2372 & 0.2903 & 0.4218\\ \midrule
\method-OPT  & \textbf{0.6882} & \textbf{0.7524} & \textbf{0.2463} & \textbf{0.3062} & \textbf{0.4464} \\
\method-OPT (w/o hierarchy) & 0.6861$^*$ & 0.7015 & 0.2374$^*$ & 0.2942$^*$ & 0.4115 \\
\method-OPT (w/o sum accumulation) & 0.6777 & 0.7436$^*$ & 0.2413 & 0.3051$^*$ & 0.4309$^*$ \\
\bottomrule
\end{tabular}
\label{tab:acc_ablation}
\end{table*}

%% file: tables/time.tex
\begin{table}[t]
\centering
\caption{Run time comparison. For each setting, the training time for one epoch is shown, as well as the performance in F1 score for the final model.}
\resizebox{\columnwidth}{!}{%
\begin{tabular}{lc|ccccc}
\toprule
 & OPT-NConcat & \multicolumn{4}{c}{\method-OPT} \\ \midrule
\# Neighbors / Fanouts & 4 & [2,2] & [8,2] & [4,4] & [2,8]\\
\midrule
Train Time (s) & 55.46 & 31.24 & 80.02 & 82.54 & 87.55 \\
Test F1 & 0.6250 & 0.6716 & 0.6835 & 0.6885 & 0.7001 \\
\bottomrule
\end{tabular}
}
\label{tab:time}
\end{table}

%% file: sections/conclusion.tex
In this paper, we introduced \method, a novel framework integrating LLMs with text-rich graph data. \method effectively leverages the text-understanding capabilities of LLMs to contextualize and compress information from rich neighborhoods in graphs. \method overcomes the inherent limitations of LLMs in processing graph data and their input length restrictions, making LLMs \textbf{adaptable} to text-rich graphs, and enabling
\textbf{effective} and \textbf{efficient}  performance on data mining tasks like node classification.

%% file: sections/appendix.tex
\subsection{The Batch Processing Algorithm }\label{app:batch_process}
In typical NLP tasks on sequential data, discrepancies in the input length of batched instances are commonly managed through padding with dummy tokens. However, in the context of graph data, this complexity is magnified due to the variability in the number of neighbors each node possesses. Although the predefined fanout parameters aim to sample equal numbers of neighbors for each node, they actually only set upper bounds, because some nodes may not have enough neighbors to reach the target fanout count.

To address these challenges, a more sophisticated padding strategy is needed to uniform both the length of the text sequence on each node and the number of neighbors of each node for all nodes in the same level of the hierarchy. As illustrated in Figure~\ref{figure:framework}, text sequences of nodes are padded with dummy tokens to the same length (e.g., $x_1$, $x_5$, $x_8$, and $x_9$ are padded to length three), and the sampled neighbors for each node are padded to match the fanout of the corresponding level (e.g., $v_2$ is padded to match fanout=3 for level one). 

Algorithm~\ref{alg:hc_batch} outlines \method for processing nodes in batches to process them efficiently. For a batch of target nodes $\V_{B} \subset \V$ indexed by $B$, lines 3 to 6 correspond to the hierarchy construction phase. In particular, line 4 samples all the nodes $\V_{B_{l}}$ that form the level $l$ for $\V_{B}$, and the nodes from the output level $L$ equal to $\V_{B}$. Line 5 gathers the node text and pads them to the same length $t$. (Note that unlike the neighbor padding that needs to be performed dynamically, individual node text padding can be performed beforehand during data pre-processing to save time, which is what we do in practice. In the pseudo-code, we explicitly list out the padding step for completeness). Lines 9 to 15 show the compression step for level $l$. To implement neighbor padding, we first initialize placeholders $\mX_{B_{l}^C}$ and $\mS_{B_{l}^C}$, which are filled with dummy tokens and have shapes according to the fanouts parameters (line 9 and line 10). Then we insert text and summary vectors corresponding to nodes from the previous level, i.e., $\mX_{B_{l-1}}$ and $\mS_{B_{l-1}}$, into the placeholder following node index $B_{l-1}$ (line 11 and line 12). Finally, the populated placeholders will be reshaped, which effectively concatenates text and summary vectors for nodes in the same neighborhood, and they are then fed to the compressor to get the summary vectors $\mS_{B_{l}}$ for nodes in $\V_{B_{l}}$ (line 13 to line 15).

\subsection{The Rearrange and Trim Algorithm}\label{app:rearrange_trim}
We show the pseudocode of the rearrange and trim function for batched token sequences in Algorithm~\ref{alg:rearrange_trim}.

\begin{algorithm}[t]
  \caption{Rearrange and Trim Batched Token Sequences}
  \label{alg:rearrange_trim}
\begin{algorithmic}[1]
  \STATE {\bfseries Input:} Token sequences $\mX$, attention masks $\mM$
  \STATE {\bfseries Output:} Compact token sequences $\tilde \mX$, Compact attention masks $\tilde \mM$
  \STATE $B$ = $\mX$.shape[0] // batch size
  \STATE $T$ = max($\mM$.sum(dim=1)) // maximum number of non-dummy tokens per sequence
  \STATE $\tilde \mX$ = A placeholder of all zeros with shape $[B, T]$
  \STATE $\tilde \mM$ = A placeholder of all zeros with shape $[B, T]$
  \FOR{$i=1$ {\bfseries to} $B$}
    \STATE $x_{selected}$ = $\vx_i$[$m_i$.bool()] // rearrange
    \STATE $T_{selected}$ = $x_{selected}$.shape[0]
    \STATE $\tilde \mX$[i, :$T_{selected}$] = $x_{selected}$ // trim
    \STATE $\tilde \mM$[i, :$T_{selected}$] = 1
  \ENDFOR
  \RETURN $\tilde \mX$, $\tilde \mM$
\end{algorithmic}
\end{algorithm}

\subsection{Detailed Experiment Settings}~\label{app:setting}
\paragraph{Hyperparameters}
Since different graphs emphasize the importance of neighbors from different hops, we consider several cases for each method. For \method and GNN, we consider three different fanouts (number of nodes to sample from neighboring nodes in each hop), 4-4, 2-8, and 8-2, and report the best result for each method in Table~\ref{tab:nc_kcore}. For NC, since the total input length of the LLM is limited (e.g., 2,048 for OPT), the LLM can usually only fit the text on the target node plus text from a few neighbors. We consider NC by sampling 3 neighbors from the 2-hop neighborhood of the target node. The full results of all settings with different hyperparameters are shown in Table~\ref{tab:nc_detailed}.

\paragraph{Implementation and Hardware}
We implement our method and LM baselines with the Hugging Face platform~\cite{wolf2020transformers}. We train these models for 20 to 60 epochs (depending on the graph size) until convergence on 8 NVIDIA A10 GPUs with 24G memory each.
We perform LoRA~\cite{hu2021lora} and gradient checkpointing to optimize GPU memory. We use DeepSpeed~\cite{rasley2020deepspeed} to perform distributed training.

\subsection{Detailed Experiment Results}~\label{app:result_detail}
In Table~\ref{tab:nc_detailed}, we report the detailed results for \method and baselines with different hyperparameters. The best result for each method is reported in Section~\ref{sec:experiment}.

\input{tables/nc_detailed}

\subsection{Experiments for Nodes in All Regions}~\label{app:result_all}
In Section~\ref{subsec:abalation}, we discussed the results of an ablation study on nodes in all regions. We now describe the detailed experiment setting and provide the detailed results. For the experiment setting, we consider a different way to split the datasets compared to the main experiment. We select all the training, validation, and testing nodes are selected randomly from the graph, with the size of each set staying the same as the main experiment (i.e. 20 nodes per class for training, 1,000 for validation, and up to 10,000 for testing). We conduct experiments with fanouts equal to [4,4] for all the GNNs and \method-OPT. Notice that the absolute performance results are not directly comparable to results in the dense regions, as the training/validation/testing nodes are all different. We show the relative performance improvement by \method-OPT for nodes in dense regions vs. all regions in Figure~\ref{figure:dense_vs_all}. Here we report the detailed performance results on nodes in all regions in Table~\ref{tab:nc_all}. 

\input{tables/nc_all}

\subsection{Experiments for Increasing Training Data}~\label{app:increase_training}
In Section~\ref{subsec:abalation}, we discussed the results of an ablation study of increasing the training data. We now describe the detailed experiment setting. Given the imbalanced label distribution in these real graphs, some classes have limited labeled samples, making it unfeasible to increase the number of training samples per class. Consequently, we have adopted a strategy of uniformly random sampling from the unused data to augment the training set. By unused data, we mean the data was not included in either train, valid, or test set in the main experiment. The label distribution of the additional training data thus should roughly follow the label distribution over the entire graph. We conduct experiments with fanouts equal to [4,4] for the OPT-GNN and \method-OPT. 

\begin{figure}[t]
\begin{center}
\includegraphics[width=\columnwidth]{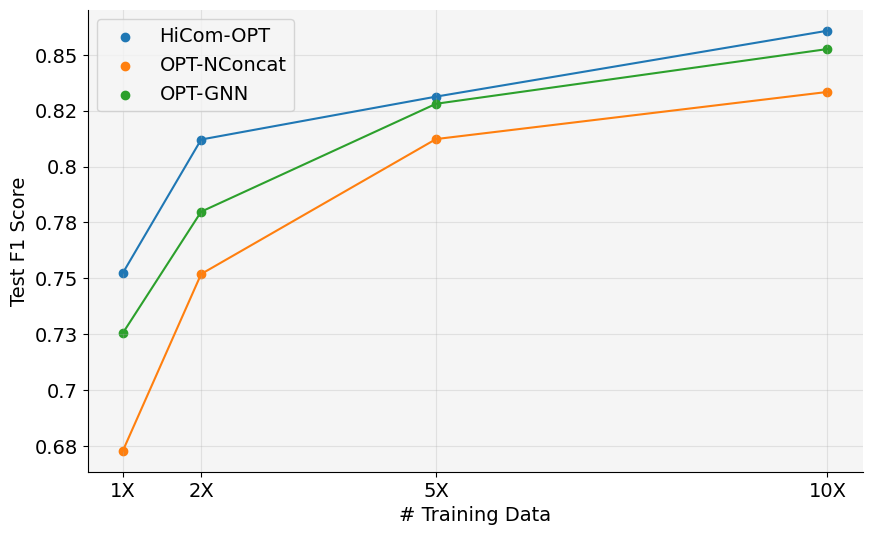}
\end{center}
\caption{Method performance on the Sports dataset with the training set in different sizes.}
\label{figure:increase_train_data_sports}
\Description{Method performance on the Sports dataset with the training set in different sizes.}
\end{figure}

We have shown the F1-score vs. training set size plot for the MAPLE graph Geology in Section~\ref{subsec:abalation}. Here we show a similar plot for an Amazon graph, e.g. Sports, in Figure~\ref{figure:increase_train_data_sports}.

 \subsection{Future Work}
The \method framework presents several intriguing avenues for future research and enhancement. A natural progression is to explore the use of more advanced LLMs, such as LLaMA~\cite{touvron2023llama}, which offer superior text-understanding capabilities. However, these models typically feature an increased number of parameters (generally $\geq7B$), necessitating substantial hardware support and optimization for efficient use within the \method framework. 
Further exploration in more tasks like link prediction or content generation based on graph structures is also exciting. Especially the generation would involve leveraging the compression and contextual understanding capabilities of \method to generate coherent and contextually relevant textual content, opening new possibilities in automated content creation. 
The adaptability and effectiveness of \method in handling text-rich graph data promise to unlock new potentials in the realm of graph-based learning and LLM applications.

%% file: tables/nc_detailed.tex
\begin{table*}[thb]
\centering
\caption{Node classification F1 score on Amazon product graphs and MAPLE citation graphs (detailed)}
\begin{tabular}{lccccc}
\toprule
Method & \textbf{\quad Cloth\quad} & \textbf{\quad Sports\quad} & \textbf{Economics} & \textbf{Mathematics} & \textbf{Geology} \\ \midrule
BERT-GNN (8-2) & 0.6136 & 0.6727 & 0.1822 & 0.2693 & 0.3667 \\
BERT-GNN (4-4) & 0.6362 & 0.6546 & 0.1730 & 0.2685 & 0.3448 \\
BERT-GNN (2-8) & 0.6262 & 0.6503 & 0.1705 & 0.2674 & 0.3405 \\  \midrule
BERT-FT-GNN (8-2) & 0.6432 & 0.6804 & 0.1656 & 0.2861 & 0.3528 \\
BERT-FT-GNN (4-4) & 0.6457 & 0.6801 & 0.1826 & 0.2677 & 0.3633 \\
BERT-FT-GNN (2-8) & 0.6455 & 0.6804 & 0.1763 & 0.2712 & 0.3529 \\
\midrule
GIANT-GNN (8-2) & 0.6267 & 0.7201 & 0.1643 & 0.2421 & 0.3074 \\
GIANT-GNN (4-4) & 0.6159 & 0.7065 & 0.1651 & 0.2366 & 0.3120 \\
GIANT-GNN (2-8) & 0.6471 & 0.6997 & 0.1599 & 0.2334 & 0.3086 \\  
\midrule
OPT-GNN (8-2)  & 0.6484 & 0.7255 & 0.1694 & 0.2634 & 0.3330 \\
OPT-GNN (4-4)  & 0.6378 & 0.6796 & 0.1747 & 0.2713 & 0.3497 \\
OPT-GNN (2-8)  & 0.6417 & 0.6607 & 0.1593 & 0.2647 & 0.3660 \\  \midrule
OPT            & 0.6541 & 0.6005 & 0.2324 & 0.2889 & 0.4206 \\
OPT-NConcat         & 0.6798 & 0.6728 & 0.2372 & 0.2903 & 0.4218 \\ \midrule
\method-OPT (8-2)   & 0.6882 & 0.6812 & 0.2463 & 0.3060 & 0.4464 \\
\method-OPT (4-4)   & 0.6622 & 0.7524 & 0.2341 & 0.2993 & 0.4339 \\
\method-OPT (2-8)   & 0.6009 & 0.7354 & 0.2331 & 0.3062 & 0.4185 \\
\bottomrule
\end{tabular}
\label{tab:nc_detailed}
\end{table*}

%% file: tables/nc_all.tex
\begin{table*}[thb]
\centering
\caption{Node classification F1 score on Amazon product graphs and MAPLE citation graphs (all regions)}
\begin{tabular}{lccccc}
\toprule
Method & \textbf{\quad Cloth\quad} & \textbf{\quad Sports\quad} & \textbf{Economics} & \textbf{Mathematics} & \textbf{Geology} \\ \midrule
BERT-GNN    & 0.5807 & 0.6054 & 0.1866 & 0.2353 & 0.3890 \\
GIANT-GNN   & 0.6549 & 0.6461 & 0.1726 & 0.2326 & 0.3662 \\
OPT-GNN     & 0.6438 & 0.6869 & 0.1707 & 0.2794 & 0.3949 \\
OPT         & 0.6537 & 0.6553 & 0.3055 & 0.3072 & 0.4654 \\
OPT-NConcat      & \emph{0.6817} & \emph{0.7154} & \emph{0.3062} & \emph{0.3131} & \emph{0.4731} \\ \midrule
\method-OPT      & \textbf{0.6849} & \textbf{0.7183} & \textbf{0.3118} & \textbf{0.3243} & \textbf{0.4737} \\
\bottomrule
\end{tabular}
\label{tab:nc_all}
\end{table*}